\def\msun{M_\odot}
\def\lsun{L_\odot}
\def\la{\mathrel{\mathpalette\fun <}}
\def\ga{\mathrel{\mathpalette\fun >}}

\def\fun#1#2{\lower0.837ex\vbox{\baselineskip0ex\lineskip0.209ex
  \ialign{$\mathsurround=0ex#1\hfil##\hfil$\crcr#2\crcr\sim\crcr}}}

\def\simpropto{\lower.2ex\hbox{$\; \buildrel \sim \over \propto \;$}}

\def\msun{M_\odot}
\def\msunyr{M_\odot \ {\rm yr}^{-1}}

\def\sles{\lower2pt\hbox{$\buildrel {\scriptstyle <}
   \over {\scriptstyle\sim}$}}

\def\sgreat{\lower2pt\hbox{$\buildrel {\scriptstyle >}
   \over {\scriptstyle\sim}$}}

\def\aprop{\mathrel{\mathpalette\fun \propto}}

\documentclass[3p,times,twocolumn]{elsarticle}
\usepackage{times}
\usepackage{lscape}

\usepackage{ecrc}

\volume{00}

\firstpage{1}

\journalname{Journal of High Energy Astrophysics}
\journal{Journal of High Energy Astrophysics}

\runauth{}

\jid{nuphbp}

\jnltitlelogo{Journal of High Energy Astrophysics}

\usepackage{amssymb}

\usepackage[figuresright]{rotating}

\begin{document}
\begin{frontmatter}




\dochead{}


\title{How \emph{Swift} is Redefining Time Domain Astronomy}


\author{N.~Gehrels}

\address{Astroparticle Physics Division, 
         NASA/Goddard Space Flight Center, 
         Greenbelt, MD 20771,
         USA}

\author{J.K.~Cannizzo}

\address{CRESST/Joint Center for Astrophysics,
          Univ. of Maryland, Baltimore County,
         Baltimore, MD 21250, USA}

\begin{abstract}
    NASA's \emph{Swift} satellite has completed ten years of
  amazing discoveries in time domain astronomy.
     Its primary mission is             to chase gamma-ray bursts (GRBs),
    but due to its scheduling flexibility it has subsequently
 become a prime discovery machine for new types of behavior.
   The list of major discoveries in GRBs and other transients includes
   the long-lived X-ray afterglows and flares
    from GRBs, the first accurate localization
   of   short GRBs,
     the discovery of GRBs at high redshift ($z>8$),
   supernova shock break-out from SN Ib,
    a jetted tidal disruption event,
    an ultra-long class of GRBs, 
   high energy emission from flare stars,
     novae and supernovae with unusual characteristics, 
            magnetars with glitches in their spin periods,
   and a short GRB with evidence of an accompanying kilonova.
          \emph{Swift} has developed a  dynamic synergism
     with ground based observatories. 
    In a few years  
     gravitational wave observatories will come on-line
      and provide exciting new transient sources for \emph{Swift} to study. 
\end{abstract}

\end{frontmatter}



\section{Introduction}
\label{}

Launched 20 Nov 2004,
\emph{Swift} (Gehrels et al. 2004)
   was  originally envisioned as primarily a 
 GRB chasing satellite.
     It has been amazing successful in this
capacity (Gehrels, 
   Ramirez-Ruiz, \& Fox 2009), but \emph{Swift}  has also evolved 
          into something much more $-$
    an all-purpose time-domain mission. 
        Its rapid response capability allows 
for   wide-ranging interactions with other observatories
    across the electromagnetic spectrum tailored to meet the demands 
              imposed by different transients.
     In a few years this will also include gravity-wave
  missions such as LIGO and VIRGO.

About once a year \emph{Swift} makes a 
    game-changing discovery.
         A partial list includes:

\smallskip

\noindent
--2005:    Short GRB mystery solution: NS-NS mergers.

\noindent
--2005:  
   Flares and bright afterglows in GRBs.

\noindent
--2008: Shock break-out in a supernova Ibc.

\noindent
--2008: Naked eye GRB from reverse shock in GRB jet.

\noindent
--2009:  Discovery of GRBs at $z>8$.

\noindent
--2010:  Galaxy  mergers in hosts of absorbed AGN.

\noindent
--2011:  Discovery of a jetted tidal disruption event.

\noindent
--2012:  SFR and metallicity evolution to $z>5$.

\noindent
--2012:  Discovery of very young (2500 year old) SNR.

\noindent
--2012:  Discovery of ultra-long class of GRB.

\noindent
--2013:  Anti-glitch in magnetar 1E 2259+586.

\noindent
--2013:  Evidence for kilonova           in a short GRB.

\noindent
--2014: Evidence for two UV color classes in SNe Ia.

\smallskip

  We will discuss these science results and others
   in time domain astronomy in this paper.
       There are many other types of steady-source science
 that \emph{Swift} addresses including surveys of
   galactic sources, AGN, and galaxies.
       In this paper, however, we cover only transient
          sources.

\section{Solution to the mystery of short GRBs}

The key to unraveling long GRBs, i.e., placing them at cosmological distances,
   came in 1997 with the first localizations and subsequent host galaxy redshift
determinations. Similarly with short GRBs the breakthrough 
     happened in May-July 2005 with afterglows
 of short GRBs discovered by 
%
  \emph{Swift} (Gehrels et al. 2004) and \emph{HETE-2}
(Ricker et al. 2003). 
  GRB 050509B was the first short GRB
  to show an 
   X-ray counterpart, 
   with a $\sim$9 arcsec localization 
   (Gehrels et al. 2005). Deep follow-up
   observations failed to reveal 
    optical afterglow 
    but did discover 
    a massive $z=0.225$ elliptical galaxy   
    near the X-ray error circle, with a
    chance coincidence probability of $\sim$$10^{-3}$ 
  (Castro-Tirado et al. 2005;
         Gehrels et al. 2005;  
           Bloom et al. 2006). 
   If GRB 050509B was indeed at $z=0.225$
   then the non-detection of  
  a supernova  was significant
    (Hjorth et al. 2005a).  
   Two months later
 \emph{HETE-2} discovered the short GRB 050709 
        (Villasenor et al. 2005).
   \emph{Chandra}
    precisely localized the X-ray afterglow 
   (Fox et al. 2005),  and 
subsequent observations revealed the first  
    optical afterglow 
     from a short GRB  (Hjorth et al. 2005b).
    The resultant
sub-arcsecond localization  placed the burst in the outer regions 
   of a star-forming galaxy at
$z = 0.160$ 
      (Fox et al. 2005). As with GRB 050509B, 
   optical follow-up also ruled out the presence of an associated supernova
(Hjorth et al. 2005b).

\begin{figure}[h!]
\begin{centering}
\includegraphics[width=3.15truein]{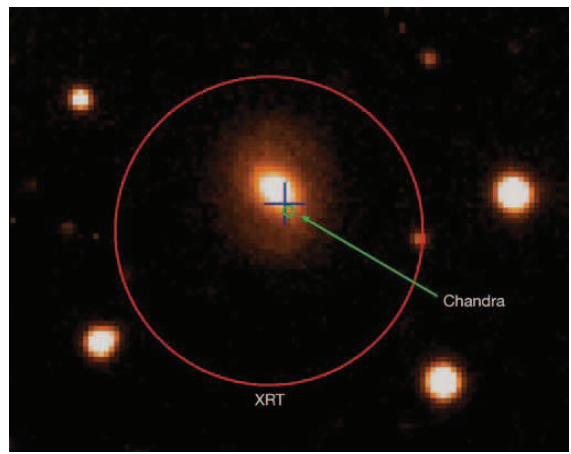}
\vskip -0.05truein
\caption{
   VLT optical image (D'Avanzo et al. 2005) 
showing the association of the short  GRB 050724
      with a galaxy (Barthelmy et al. 2005). 
   Blue cross is the position of the 
  optical transient
    (Gal-Yam et al. 2005; D'Avanzo et al. 2005).
\emph{Swift}/XRT (red circle) and \emph{Chandra} 
   (green circle) burst positions are
superimposed on a bright red galaxy
  $z=0.258$ (Prochaska et al.
   2005), 
 implying a
low-redshift elliptical galaxy as the host. 
  The XRT position has been 
         revised from 
                Antonelli et al. (2005) 
 by astrometric comparison with objects
in the field. The projected offset 
       from the center of the galaxy corresponds to
$\sim4$ kpc assuming the standard cosmology.
    }
\label{fig:scott}   
\end{centering}
 \end{figure}

  The final triumph of 2005 for short GRBs was another \emph{Swift}
    discovery, the short GRB 050724  (Barthelmy et al. 2005) 
    which resulted in
   the discovery of X-ray, optical/near-IR, and the 
    first radio afterglow (Figure \ref{fig:scott}).
          It yielded another 
  subarcsecond   
   localization in an
elliptical galaxy, at $z = 0.257$ 
             (Berger et al. 2005b). 
   The    ratio    of radio to X-ray afterglow
emission also demonstrated that both the energy and density 
     scale were lower than for long GRBs (Berger
et al. 2005b). Taken together, these three early events 
            showed that short GRBs are cosmological, 
    they produce afterglow emission similar 
    to that of long GRBs but with a lower energy and
density scale, and 
  their progenitors are not massive stars (given their lack of
       associated supernovae).

Ten years after the discovery of short GRB afterglows more than 70 short GRBs
      have been found by \emph{Swift} and other
   $\gamma-$ray satellites (Berger 2014).
              A sizable fraction have X-ray and optical
  afterglows; a few have been detected in the radio.
%
%
  The localizations and optical follow-up work
     have identified $\sim$40 host galaxies and enabled
     detailed studies of the intragalactic locations
  of short GRBs.
%
   An 
  \emph{HST} study  
  of 10 short GRBs
   within their host galaxies 
   reveals 
   they 
     trace the light distribution of their hosts, while
   long GRBs are concentrated in the brightest regions
     of their host galaxies (Berger et al. 2005a;
                       Fong, Berger, \& Fox 2010).

      Fong et al. (2010)  find that 
 the median value of the projected offset from host center for 
      short GRBs of $\sim$5 kpc is about 5 times larger than that for
           the corresponding long GRB median offset. 
                 Interestingly,
    when the two offset distributions are normalized to the size of
 the host galaxy, they lie almost on top of each other. 
              In other words, the
  host galaxies for long GRBs are 
       $\sim$1/5  
          as large on average
             than those of short GRBs.
   In an updated \emph{HST}
      study using 22 short GRBs Fong \& Berger 
  (2013) 
     refine their previous results, and furthermore
   find that short GRBs strongly under-represent their
    hosts' rest-frame optical and UV light; a fraction
  $\sim$$0.3-0.45$  are located in regions with no stellar light,
   and $\sim$0.55 in regions with no UV light.
      Therefore Fong \& Berger conclude that short GRB progenitors
    must migrate over considerable distances before their eventual explosions,
   which supports the idea of progenitor 
    kicks in  compact binary systems
   and  the NS-NS merger model for short GRBs.
%

\section{Flares and bright afterglows in GRBs}


\begin{figure}[h!]
\begin{centering}
\includegraphics[width=3.15truein]{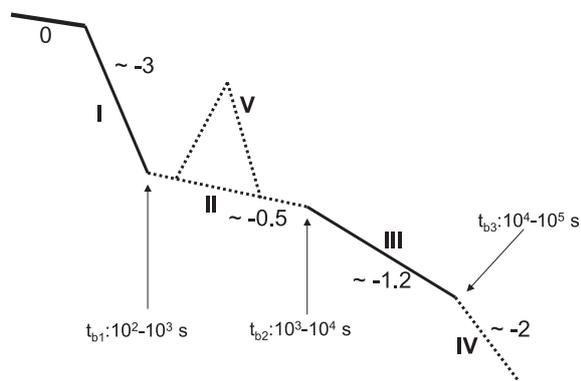}
\vskip -0.05truein
\caption{
Schematic representation on a $\log f_X -\log t$ scale of the
  now canonical X-ray decay light curve for GRBs
    based on observational data from \emph{Swift}/XRT
  (Zhang et al. 2006).
  The transition times in seconds after $T0$ are given,
  as well as the approximate power law decay indices
   associated with each segment.
    Phase ``0'' corresponds to the prompt emission.
  Sometimes bright flares accompany the plateau phase, segment II.
    }
\label{fig:bingz}   
\end{centering}
 \end{figure}

%
      By the time \emph{Swift} had discovered a handful 
  of GRBs and their XRT
      X-ray light curves were studied in detail,
  several basic features 
   became obvious 
     (Zhang et al. 2006; 
     Nousek et al. 2006;
    O'Brien et al. 2006). 
   This now standard picture of the X-ray
      decay came as somewhat of a surprise and certainly
       was not expected before \emph{Swift}. 
%

%
%
   The canonical X-ray behavior, depicted schematically in
   Figure \ref{fig:bingz},  
   consists 
of up to  five  distinct          $f_X\aprop t^{\ \alpha}$ segments: 
   (0) emission coincident with the prompt emission,
   (I) an initial  
steep decay $-5 \la \alpha \la -3$,  
  (II) a  
shallow decay 
   $-1 \la \alpha  \la -0.5$, 
 (III) a steeper 
decay  
   $-1.5 \la \alpha \la -1$, 
and finally
  (IV)  a slightly steeper decay
   $ \alpha \simeq -2 $.
 The transition times between segments are given in Figure \ref{fig:bingz}.
  In addition, one often sees (V) large flares superposed on the shallow
    decay branch. 
%

    The initial steep decay 
(Tagliaferri et al. 2005)
    is associated with
       the tail   
          of the prompt emission.
      Early speculation was
   that it is due to 
          photons
      that are radiated at large angles relative to our line
  of sight in the initial fireball
    -- so called ``high-latitude'' emission 
     (Kumar \& Panaitescu 2000).
 However, subsequent study showed that the predicted relation
       between spectra index and temporal decay index was not borne out   by the   
         data (O'Brien et al. 2006, see their Fig. 4).
 
    The prolonged emission associated with the shallow decay
  prompted much theoretical work on ``continuous'' or ``delayed''
    energy injection models.
            It was thought that
  the forward shock keeps 
   being refreshed for some time,
  possibly due to 
  a long-lived central engine,  a wide 
  distribution of Lorentz factors in the jet, or 
   possibly the deceleration of a 
  Poynting flux-dominated 
  jet. These speculations led to a variety 
  of models based on fall-back accretion
            disks 
   (Kumar, Narayan, \& Johnson 2008; 
       Cannizzo \& Gehrels 2009),
magnetars 
         (Dall'Osso et al. 2011;
      O'Brien \& Rowlinson 2012;
          Rowlinson et al. 2013),
              and various other ideas
          such as ``prior emission''
     (Yamazaki 2009).

Recent work based on  
  relativistic hydrodynamical computations
    favors the standard 
picture in which a baryon dominated
    jet directed along our line-of-sight
             breaks through the photosphere of a progenitor
  (Duffell \& MacFadyen 2014).
  This model
      may be able to account for all the decay phases observed 
  by XRT   (Duffell 
    \& MacFadyen 2014).
  In this (now canonical)
     model a jet launched ballistically from near the
         progenitor core with a bulk Lorentz factor
  $\Gamma_{\rm bulk}
   \simeq20-30$ undergoes strong shock heating and lateral
        confinement, so that when it emerges from the 
             progenitor it has a large
                    internal energy 
  $\Gamma_{\rm thermal} \sim p/\rho  \simeq 10$.
    The subsequent jet expansion 
  due to the large $\Gamma_{\rm thermal}$
         leads to an effective Lorentz factor 
  $\Gamma_{\rm eff}  \simeq 2 \Gamma_{\rm bulk}
   \Gamma_{\rm thermal} \simeq400-600$  
                  for a distant, line-of-sight observer.
The GRB prompt $\gamma-$radiation 
   is 
     produced by strong internal
      shocks in the expanding fireball 
         at the point where 
           it becomes optically thin to its own radiation.

\begin{figure}[h!]
\begin{centering}
\includegraphics[width=3.05truein]{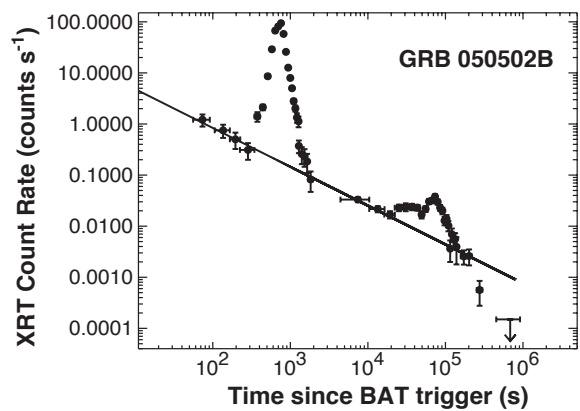}
\vskip -0.05truein
\caption{ \emph{Swift}/XRT 
     X-ray light curves of the afterglows of GRB 050502B
   (Burrows et al. 2005).
   The best fit decay, shown by the solid line, is 
  $\alpha = -0.8 \pm 0.2$.
  The bright X-ray flare is superposed
         on this underlying power-law decay.
   At later times the    
  light curve    
   has several bumps, both suggesting
  late-time energy injection into the external shock
    or continued internal shock activity. The rapid decline in
    count rate for GRB 050502B at $t >  10^5$ s indicates a
     possible jet break at $\sim$$1-2$ d postburst.
    }
\label{fig:flare}   
\end{centering}
 \end{figure}

Large
X-ray flares associated
   with GRBs (Chincarini et al. 2007; 
               Falcone et al. 2007)
   suggest that the GRB central engine 
  is very likely still active after the prompt gamma-ray emission 
  is over, but with a reduced activity at later times. 
  Figure \ref{fig:flare}  
  shows a giant flare in GRB 050502B
   with a    
  total energy  comparable to the burst itself.
   These strong, rapid X-ray flares have been seen in other bursts
  and indicate  
  that the central engines of the bursts have
    long periods of activity, with strong internal 
  shocks continuing for hundreds of seconds after 
  the gamma-ray emission has ended.
   Flares have also been seen in short GRBs, e.g. GRB050724
  (Grupe et al. 2006).

\section{GRBs and Supernovae}

On 18 February 2006 \emph{Swift}
          detected the remarkable burst GRB 060218 
   which shed light 
        on the connection between SNe and GRBs.
It lasted longer than and was softer than any previous burst, and was associated
with SN 2006aj at only $z=0.033$. The BAT trigger 
        enabled XRT and UVOT
observations during the prompt phase of the GRB and initiated multiwavelength
observations of the supernova from 
    the time of initial core collapse. The 
spectral peak in prompt emission at $\sim$5 keV 
   places GRB 060218 in the X-ray flash
category of GRBs (Campana et al. 2006), the 
   first such association for a GRB-SN
event. Combined BAT-XRT-UVOT observations provided the first direct
observation of shock-breakout in a SN 
          (Campana et al. 2006). This is inferred from
the evolution of a soft thermal component in the X-ray and UV spectra, and early
time luminosity variations. 
     SN 2006aj was dimmer by
a factor $\sim$2 than previous SNe associated 
   with GRBs, but still $\sim$$2-3$ times
brighter than normal SN Ic not associated 
   with GRBs (Pian et al. 2006; Mazzali et al. 2006).
       GRB 060218 was an underluminous burst, as were two of the
other three previous cases. Because of the low luminosity, these events are only
detected when nearby and are therefore rare occurrences. However, they are 
actually $\sim$$5-10$ times more common in the universe 
 than normal GRBs (Soderberg
et al. 2006). 
  For completeness we note there have also 
     been  two nearby long GRBs 
 with no associated SNe to deep limits
   (Gehrels et al. 2006;
      Fynbo et al. 2006;   
    Della Valle et al. 2006).

\emph{Swift} has added to the list of GRB-with-associated-SNe, 
   most notably with GRB 130427 
          (Maselli et al. 2014; 
          Melandri et al. 2014;
            Levan et al. 2014).
   Its energy  $E_{\rm iso} \sim 10^{54}$ erg
       ranks it among the most luminous 5\% of GRBs observed 
   by \emph{Swift},
     and a factor of 100 brighter than GRB 030329
     (Hjorth et al. 2003)
    which was the previously most luminous GRB with a well-studied SN.
        Its low redshift $z=0.340$ (Levan et al. 2013; 
  Xu et al. 2013; Perley et al. 2014) 
     means that it is readily accessible
      for spectroscopic
study of an attendant SN, and indeed one has been seen -- 
   SN 2013cq
  (de Ugarte Postigo et al. 2013; Xu et al. 2013).

\begin{figure}[h!]
\begin{centering}
\includegraphics[width=3.15truein]{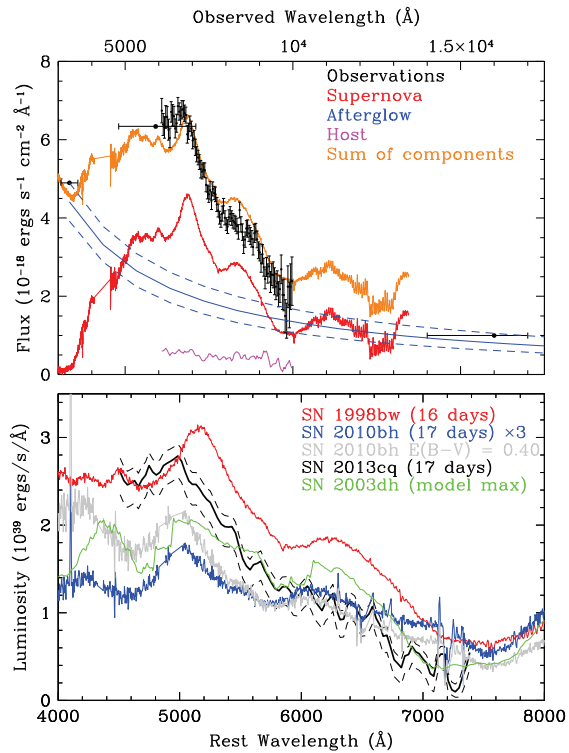}
\vskip -0.005truein
\caption{ 
Spectral energy distribution of GRB 130427A/SN 2013cq as
measured with \emph{HST} (Levan et al. 2014). 
   Top panel shows the data (black) along with the different
components that may contribute as indicated. The host galaxy spectrum is based
on  an extraction of the host directly under the GRB position, and not its global
properties. The lower panel shows the smoothed SN spectrum after subtraction
of the afterglow light, and in luminosity space, directly compared with spectra
of other GRB/SNe pairs. The supernovae have been scaled as shown in the
legend, but in general the spectra show a good match with SN 1998bw at a
similar epoch.
\label{fig:levan}   
    }
\end{centering}
 \end{figure}

Levan et al. (2014)
   present \emph{HST} 
observations of
GRB 130427A (Figure \ref{fig:levan})
             which show that it
was associated with a luminous broad line SN Ic (SN 2013cq).
The red spectra offer good agreement with those of SN 1998bw,
while the bluer spectra appear well-matched in position and 
shape to   SN 2010bh. 
  Levan et al. 
   utilize \emph{HST}
   to separate the afterglow, host, and SN contributions to the observed light
 at  $t\sim$17 d (rest frame).
   ACS grism observations show that the associated supernova,
  SN 2013cq, has an
   overall spectral shape and luminosity similar to SN 1998bw
   (with a photospheric velocity $v_{\rm ph} \simeq 0.05c$).
        The blue features in the spectrum are matched better 
    by the higher velocity SN 2010bh ($v_{\rm ph} \simeq 0.1c$),
      however SN 2013cq is significantly fainter.
           The burst originated $\sim$4 kpc from the 
   nucleus of a moderately star-forming region,
  $\sim$$1\msunyr$. The absolute magnitude, physical size, and morphology of 
  the host galaxy, as well as the location of the GRB within it, are 
    very similar to those of GRB 980425/SN 1998bw.
 The similarity of the SNe and environment from both the most
   luminous and least luminous GRBs suggests that
broadly similar progenitor stars can create GRBs across 
  six orders of magnitude in isotropic energy.

\section{SN 2008D Shock Breakout}

\begin{figure}[h!]
\begin{centering}
\includegraphics[width=3.20truein]{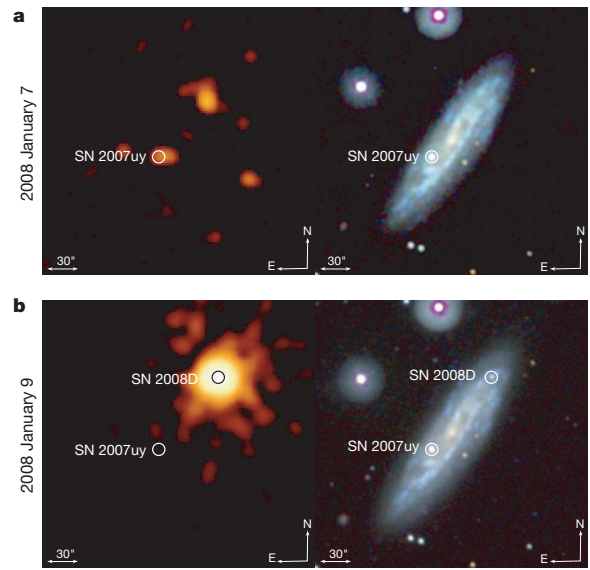}
\vskip -0.35 truein
\caption{
   X-ray (left) and optical (right) discovery images for  
     SN2008D
   (Soderberg et al. 2008).
       }
\label{fig:soderberg}   
\end{centering}
\end{figure}

The $t=0$ time of a SN is marked by a burst of 
 neutrinos, thus the ``delayed'' optical light from
radioactivity in the ejecta through which most SNe 
  are discovered does not provide information
about the first moments following the explosion. 
   On 9 January 2008 \emph{Swift}/XRT serendipitously
discovered an extremely bright X-ray transient (Figure \ref{fig:soderberg})
 while carrying out 
a preplanned observation of the 
  nearby ($d=27$ Mpc) 
     galaxy NGC 2770 
    (Soderberg et al. 2008). 
    Two days earlier XRT had
observed the same location and did not see a source. 
  X-ray outburst (XRO) 080109 lasted about
400 s and occurred in one of the galaxy's spiral arms. 
   XRO 080109 was not a GRB (no $\gamma-$rays were
detected), and the total X-ray energy $E_x \simeq  
    2 \times 10^{46}$ erg was orders of magnitude lower than a GRB.
The peak luminosity $\sim$$6 \times 
      10^{43}$ erg s$^{-1}$ is much 
      greater than the Eddington luminosity for
    a $\sim$$1\msun$
object, and also from type I X-ray bursts. 
   Therefore the standard accretion and thermonuclear ﬂash
scenarios are excluded.

Simultaneous  \emph{Swift}/UVOT observations did not reveal 
  a counterpart, but UVOT observations
at 1.4 hr showed a brightening. Gemini-North 
  observations beginning at 1.7 d 
revealed a spectrum suggestive of a young SN 
    (Soderberg et al. 2008). Later observations confirmed
the spectral features. The transient was classified as a 
  type Ibc SN based on the lack of H, and weak
Si features.

Soderberg et al. (2008) argue that the X-ray flash 
  indicates a trans-relativistic shock breakout
from a SN, where the radius at breakout is $> 7
    \times 10^{11}$ cm, and the shock velocity at breakout 
 $\beta\gamma \la  1.1$. 
  They                      estimate a circumstellar density which yields an inferred pre-SN
mass loss rate $\sim$$10^{-5} \msunyr$, 
         reinforcing 
  the notion of a Wolf-Rayet progenitor. The similarity
between the shock break-out properties of the 
          He-rich SN 2008D and the He-poor GRB-associated
SN 2006aj are consistent with a dense stellar wind around a compact Wolf-Rayet progenitor.

X-ray and radio observations presented by 
        Soderberg et al. (2008) of SN 2008D are the earliest
ever obtained for a normal type Ibc SN. At $t<10$ d, 
     the X-ray and peak radio luminosities are orders
of magnitude less than those of GRB afterglows, but comparable to those of normal type Ibc SN.

Mazzali et al. (2008)
  highlight several unusual features associated
with SN 2008D/XRF 080109:
   (i) a weak x-ray flash (XRF), 
   (ii) an early, narrow optical peak, 
  (iii) the
       disappearance of the broad lines 
    characteristic of SN Ic HNe, 
   and 
   (iv) the development of He lines as
in SNe Ib. By analyzing its light curve  Mazzali et al.
  infer a SN energy $\sim$$6\times 10^{51}$ erg
 and ejected mass $\sim$$7\msun$,
   placing it between normal SNe Ibc and HNe.
    Mazzali et al. conclude that SN 2008D
  was among the weakest explosions producing
   relativistic jets,  in accordance with the inference
of Soderberg et al. of a trans-relativistic shock breakout.

 \section{ \emph{Swift} reverse shock, naked eye GRB}




   In the standard fireball
model,  
   relativistic shells within a jet
   propagate away from the 
central engine and  
      into the surrounding medium,
      generating a forward shock (FS).
   A reverse shock (RS) 
  propagates back into the jet.
   Studies of the GRB afterglow FS/RS emission
        can potentially  provide information
   about the explosion energy, geometry, and structure
  of the circumburst medium
     (Sari, Piran, \& Narayan 1998;
       Chevalier \& Li 2000). 
       The 
   most useful probe of the initial bulk Lorentz factor 
   $\Gamma_{\rm eff}$  and the
       ejecta composition is the RS.
   The combination
      of large RS speed $v_{\rm RS} \sim c$
         and the finite and limited 
  ejecta length  
     means that the 
   only hope of directly observing 
  the RS  
    is via its early-time emission,
  basically  optical and/or  radio detections.
   To be detectable 
            very bright bursts are needed.
   The RS emission 
is expected to produce a synchrotron spectrum 
   similar to the FS, with well-defined
   RS/FS properties
    (Sari \& Piran 1999ab; 
    Kobayashi \& Zhang 2003ab; 
      Zou et al. 2005).
     There have been detections of hints of an RS-like 
    component in a handful of bursts.
%
   A detailed understanding of RS emission
requires a careful decomposition of the afterglow spectral energy
distribution (SED) into RS and FS components. 
  Since the peak
frequencies of the two components are related by a factor of
${\Gamma_{\rm eff}}^2 \ga 10^4$, 
     such a decomposition requires multi-wavelength
observations spanning several orders of 
   magnitude in frequency (Laskar et al. 2013).



\begin{figure}[h!]
\begin{centering}
\includegraphics[width=3.115truein]{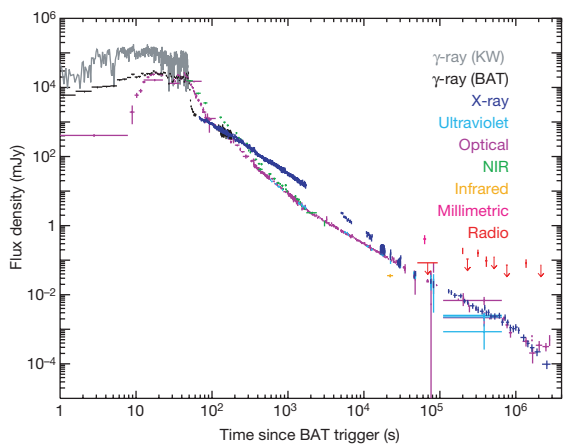}
\vskip  -0.05truein
\caption{
   Broadband light curve of the naked eye GRB 080319B,
including radio, mm, IR, NIR, optical, UV, X-ray and
$\gamma-$ray flux densities (Racusin et al. 2008). 
   The UV, optical and NIR data are normalized to
the UVOT $v-$band 
  for $ 1.5 \ {\rm ks} \  < (t-T0) <  10 \ {\rm ks}$. 
\emph{Swift}/BAT data are extrapolated 
   down into the XRT bandpass ($0.3-10$ keV)
for direct comparison with the XRT data.
        Combined X-ray and BAT data
are scaled up by a factor of 45, and the 
 \emph{Konus-Wind} data are scaled up
by a factor of $10^4$ for comparison 
     with the optical flux densities. This figure
  includes 
   one VLA radio data point
  (Soderberg, Chandra, \& Frail 2008),  
    and optical data
from     KAIT, Nickel and Gemini-South 
        (Bloom et al. 2009).
    }
\label{fig:racusin}   
\end{centering}
 \end{figure}

On 19 March 2008 \emph{Swift} 
  detected the naked eye GRB 080319B (Racusin et al. 2008).
    It was the brightest optical burst ever
   observed.
 If it were 2 kpc from Earth it would have been as 
  bright as the noon  sun in the sky.
       It had a redshift $z=0.937$, a peak visual
  magnitude 5.3, and a total energy 
    in $\gamma-$rays 
        $E_{\rm iso}=
          1.3\times10^{54}$ erg (20 keV -- 7 MeV).
  This burst  (Figure \ref{fig:racusin})
  was observed with a wide variety of instruments
    spanning the spectrum from radio to $\gamma-$ray.
   The earliest data at $t < 50$ s  reveal a common shape 
      for the bright optical and $\gamma-$ray light curves,  
             indicating they arise from
  the same physical region.
   The second optical component (50 s $<t<$ 800 s) shows the
   distinct characteristic of
     a RS, namely, an excess 
      above a time-reversed extrapolation 
from the later optical power law decay.
    The final component
(at $t>800$ s) is the afterglow produced as the external 
FS  propagates into the surrounding medium.
   Previous 
measurements of GRBs had  
    never revealed all three optical components
      in the same burst with such clarity.
%
%
%
%
%
 
  RS emission  
  did not become visible 
  until the  prompt emission faded.
    The
high peak luminosity of the optical RS so soon
after the end of the $\gamma-$ray emission 
        indicated that the RS 
was at least mildly relativistic.
   Furthermore, the GRB outflow could not have been
   highly magnetized 
when it crossed the RS or the 
   RS itself would have been suppressed.
       On the other hand, the presence of strong
 optical emission accompanying the RS demands
    some magnetization, therefore an intermediate magnetization
  seems to be indicated (Racusin et al. 2008).

\section{GRBs at $z > 8$}

GRBs
    can serve as powerful 
     probes of the early universe.  
 GRB afterglows have intrinsically very simple spectra
   thereby allowing robust redshifts
from low signal to noise spectroscopy, or photometry. 
During a fortuitous one week span in April 2009
   \emph{Swift}
    found two GRBs with redshifts greater than eight.
          These are two of the most distant objects
     ever found;  
           they harken back to a time when the 
     universe was only $\sim$0.5 Gyr old.  A study of the light from these
     ancient beacons can provide crucial clues
     about the early universe.        
     Their luminous afterglows reveal locations and
    properties of star forming
  galaxies at distant redshifts,
 potentially localizing first generation (Pop III) stars.

\begin{figure}[h!]
\begin{centering}
\includegraphics[width=3.20truein]{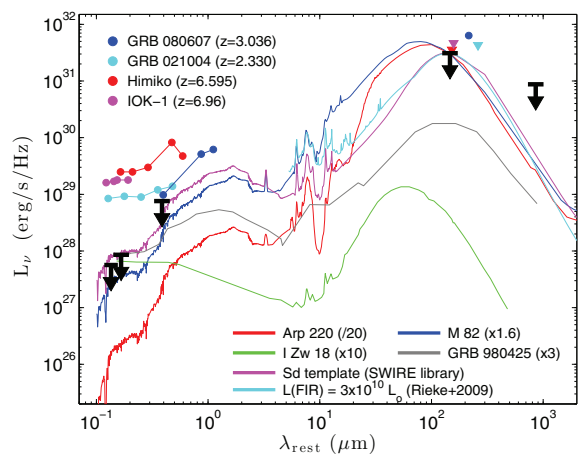}
\vskip -0.05truein
\caption{
               Limits on the rest-frame
            luminosity density, as a function of rest-frame wavelength, of the host
         galaxy of GRB 090423 (Berger et al. 2014)
      in the near-IR (\emph{HST}), mid-IR (\emph{Spitzer}), mm (ALMA),
  and radio (ATCA). Also plotted: SEDs of the ULIRG Arp 220 (red), the starburst
   galaxy M82 (blue), the dwarf I Zw 18 (green), the host galaxy of the
   other $z > 8$ burst,  GRB 980425 (gray, from Micha\l{}owski et al. 2014),
  an Sd galaxy template (magenta), and a template for a galaxy with
  $L_{\rm IR} = 3 \times 10^{10}L_{\odot}$ (cyan, from Rieke et al. 2009), all shifted
  to $z=8.23$ except for I Zw 18 and the host galaxy of GRB 980425, which are scaled
  to the \emph{HST} limits.   
  Also shown are ALMA observations and rest-frame UV/optical SEDs
     of two other GRB host galaxies (circles: detections;
triangles: upper limits;
   GRB 080607 is a marginal $3.4\sigma$ detection;
      Wang et al. 2012), and two spectroscopically confirmed
   Ly$\alpha$ emitters at $z \approx 6.6-7.0$ (Ouchi et al. 2013;
Ota et al. 2014).
    }
\label{fig:berger_limits}   
\end{centering}
 \end{figure}

{\bf GRB 090423:}
  Tanvir et al. (2009) present observations of GRB 090423
     taken 
   with a variety of instruments, including XRT, UKIRT, and VLT.
    An SED at $\sim$17h gives a photometric redshift $z\approx 8.1$.
      VLT observations starting at $\sim$17.5h
 show a flat continuum disappearing at $\lambda \la 1.13 \mu$m,
        which  confirms the origin of the break
     as Ly$-\alpha$ absorption
  by neutral H at $z\approx 8.2$.
   A simultaneous best fit to both spectra and photometric
     data points gives $z=8.23^{+0.06}_{-0.07}$.
   Salvaterra et al. (2009) compare 
    rest-frame $\gamma-$ray and X-ray light
curves of GRB 090423 with those of GRBs at low, intermediate, and
    high redshifts and find them to be remarkably similar.
  
 Far-IR 
    observations of the host galaxy of GRB 090423 
  with ALMA and \emph{Spitzer} 
   (Figure \ref{fig:berger_limits})
   taken by  Berger et al. (2014) reveal that
     the host is not seen
 at rest frame wavelengths of 145 $\mu$m (ALMA)
  and 0.39 $\mu$m (\emph{Spitzer}),
     with  inferred upper limit 
        $L_{\rm IR} \la 3 \times 10^{10} \lsun$.
  This corresponds to 
  an obscured star formation rate $\la 5\msunyr$.
 %
  The \emph{Spitzer} and \emph{HST} upper
     limits place a limit on the host galaxy
  stellar mass $\la 5\times 10^7 \msun$ 
        (assuming 
         a $100$ Myr old stellar population with constant
         star formation rate).

\begin{figure}[h!]
\begin{centering}
\includegraphics[width=3.30truein]{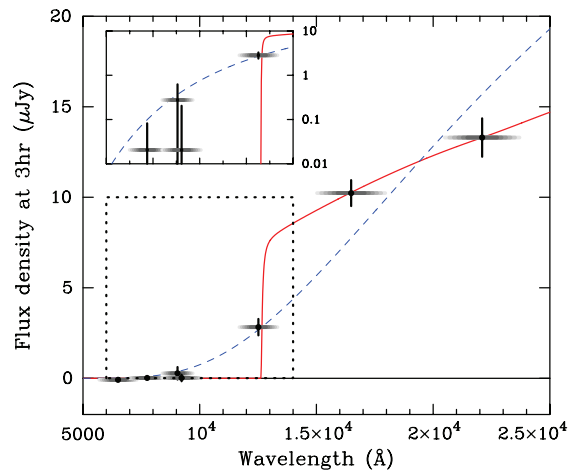}
\vskip -0.05 truein
\caption{
       Spectral energy distribution of the GRB 090429B afterglow  
           (Cucchiara et al. 2011).
    Horizontal shaded bars
illustrate the widths of the broadband filters. 
 Solid red line indicates the best fit model  
   ($\chi^2/dof = 1.76/3$), 
   with fitted parameters 
    $z = 9.36$, rest-frame extinction 
      $A_V = 0.10$, and intrinsic power-law
slope $\beta_O  = 0.51$. The inset shows the short wavelength
   region 
(indicated by a dotted box) on a log scale to show
     more clearly
 the constraints from the optical measurements. An alternative low-redshift
($z \approx 0$), high-extinction ($A_V = 10.6$) model
   is shown as a dashed blue line, but
in fact is formally ruled out at high significance
      ($\chi^2/dof = 26.2/4$).
   }
\label{fig:cucchiara}   
\end{centering}
\end{figure}


{\bf GRB 090429B:}
  Cucchiara et al. (2011)  infer  a photometric
redshift  $z \simeq 9.4$   
   based on deep observations
with Gemini,  VLT, and GROND (Figure \ref{fig:cucchiara}).
   The 90\% likelihood range is $9.02 < z < 9.50$, and the lowest
redshift allowed at 99\% confidence is $z > 7.7$. 
  The non-detection of the host galaxy
to deep limits ($Y[AB]\sim28$,  or $\sim$$0.001L*$ at $z=1$
      where $L*$ is the
    characteristic galaxy luminosity)
       in late time 
 \emph{HST} observations
strongly supports the extreme redshift since 
  \emph{HST}  would have   
   detected any low$-z$ galaxy, 
   even if it were extremely dusty. 
   Finally, the energetics of
GRB 090429B are comparable to those of other GRBs, and suggest that the progenitor
of GRB 090429B is not greatly different 
    from those of lower redshift bursts.
%

\section{ GRB studies using complete samples}

  As the total number of \emph{Swift} GRBs approaches 1000,  it becomes not
only possible but imperative to carry out studies based on well-defined
  and complete samples.  Such samples are essential in our next steps in
  understanding GRBs and using them as probes of the universe while at the
  same time controlling for and minimizing the effects of observational bias.
     A major step in this direction have been studies based on the ``TOUGH''
   survey -- The Optically Unbiased GRB Host Survey
          (Hjorth et al. 2012; 
        Jakobsson et al. 2012;
  Milvang-Jensen  et al. 2012;
         Kr\"uhler et al. 2012;
   Micha\l{}owski et al. 2012).

   Hjorth et al. (2012) define a homogeneous subsample of 69 \emph{Swift} GRB galaxies
   using well-defined criteria aimed at making the sample optically
  unbiased. Using VLT they detect host galaxies for 80\% of the GRBs in the sample.
  For those hosts  with redshifts, 38 in all,  they determine a median value $2.14\pm0.18$.
     Jakobsson et al. (2012) increase the total number of TOUGH redshifts from 38/69 to 
    53/69, spanning a range $0.345 \la z \la 2.54$. They constrain
  the fraction of \emph{Swift} GRBs to a maximum of 14\% for $z>6$ and 5\% for $z>7$.
     Milvang-Jensen et al. (2012) search for Ly$-\alpha$ emission in a subsample of 20 host galaxies.
  They detect Ly$-\alpha$ emission from 7 of the 20, with luminosities in the range 
   $(0.6-2.3)\times 10^{42}$ erg s$^{-1}$. 
  Kr\"uhler et al. (2012) use NIR spectroscopy to refine the $z$ database, and 
   Micha\l{}owski et al. (2012) use radio observations to infer that TOUGH galaxies
   are similar in many respects to other star-forming galaxies at $z \la 1$.  
 
      Kohn et al. (2015) present an analysis of FIR 
  properties
     of an unbiased sample of 20 
  \emph{BeppoSAX} and \emph{Swift}
   host galaxies at $<z>=3.1$
   and conclude that the detection rate of
  GRB hosts is consistent with that predicted
   assuming GRBs trace the cosmic SFR in 
 an unbiased way.

%
%
%
%
%


Similar efforts have been undertaken based on carefully selected
sub-samples of GRBs, characterized by a high degree of completeness
     in redshift determination,
             which are bright in the $15-150$ keV BAT band
  (Salvaterra et al. 2012;
         Nava et al. 2012;
     Campana  et al. 2012;
     D'Avanzo et al. 2012;
    Ghirlanda et al. 2013).
 Ghirlanda et al. (2013)
   construct a homogeneous sub-sample
 of 38 radio detections/upper limits
  from a complete sample of 58 bright
 \emph{Swift} long GRBs.
    They find that GRBs which typically
trigger \emph{Swift}/BAT can be
detected by JVLA within a few days
with modest exposures, even
at high $z$.

\section{           Swift J1644+57 $-$ the 
   first Jetted Tidal Disruption Event}




Tidal disruption events (TDEs)
   are caused by the tidal disruption of
stars that venture too close 
    to the massive black holes (MBHs)
at the centers of galaxies 
          (Rees 1988; 
        Phinney 1989;
        Cannizzo, Lee, \& Goodman 1990).
   Prior to March 2011, nearly all our observational
   information was based on optical/UV studies
  (Gezari et al. 2006, 2008)
    or
  long-term X-ray data with poor time sampling (Komossa et al. 2004).
  This changed with the discovery by \emph{Swift}
 of GRB 110328A/Swift J1644+57,
  a TDE viewed down the jet axis of a MBH
in the nucleus of a galaxy
  at $z=0.354$ 
       (Bloom et al. 2011;
      Burrows et al. 2011; 
        Levan et al. 2011;
       Berger et al. 2012).

\begin{figure}[h!]
\begin{centering}
\includegraphics[width=3.15truein]{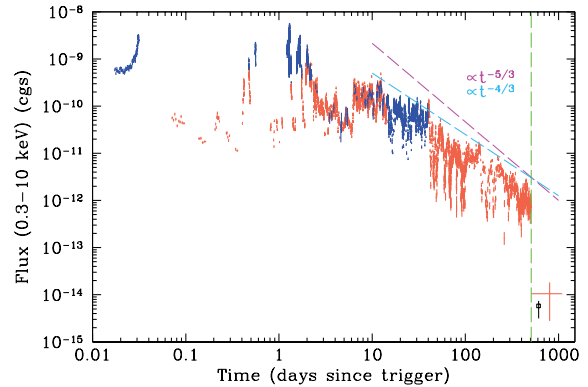}
\vskip -0.05truein
\caption{ The long term XRT light curve for
  Swift J1644+57, the jetted tidal 
     disruption event (Mangano et al. 2015).
The decay closely follows $-4/3$; a decay $-5/3$ is disfavored.
     }
\label{fig:vanessa}   
\end{centering}
 \end{figure}

A TDE occurs when the radius of closest approach
  $R_P$ of a star passes within the tidal disruption radius
  $R_T$.
After a TDE occurs, 
   there is no accretion of shredded
   stellar material onto the SMBH  for a time $\sim$$t_{\rm fb}$,
   the fallback time for the most tightly
  bound debris. 
Therefore, one expects  a gap of $\sim$$t_{\rm fb}$,
  after which accretion can begin. In addition,
   Tchekhovskoy et al. (2014) point out there will also
     be an additional $\Delta t_{\rm offset}$ 
     after accretion starts  before
 the jet activates. Thus one anticipates a total time
  interval $t_{\rm int} \equiv t_{\rm fb} + \Delta t_{\rm offset}$
    between TDE and an observed jet activity,
  i.e.,  flaring followed by a  
     decay $\propto (t/t_{\rm int})^{ \alpha}$.
Thus in an idealization in which 
       (i) we view the TDE down the jet axis,
          (ii) the jet power tracks the rate of
                accretion onto the SMBH, and 
                (iii) most of the jet power comes out in X-rays,
\begin{equation}
\ f_{X} = \left\{
   \begin{array}{l l}
             0   &   \quad  {  t  < t_{\rm int} }\\
    f_{\rm X, max}(t/t_{\rm int})^{\alpha}
                & \quad {  t\ge t_{\rm int} }
    \end{array}  \right.\
\end{equation}
    By considering the ratio of the peak X-ray flux to the fluence
$\Delta E_{X} = \int_{t_{\rm int}}^{\infty} f_{X}(t) dt$,
   one can directly measure                   $t_{\rm fb} + \Delta t_{\rm offset}$. 
     From the functional form for $f_{X}(t)$ one may write
\begin{equation}
   t_{\rm fb} + \Delta t_{\rm offset} 
               =  -(1+\alpha) {\Delta E_{X} \over f_{\rm X, max}}.
\end{equation}
      Note that all uncertainties such as beaming angle,
    accretion efficiency, jet efficiency, etc., cancel out. 
      Using the \emph{Swift}/XRT 
  measured values  $f_{\rm X,  max} \simeq 9 \times 10^{-9}$ 
    erg cm$^{-2}$ s$^{-1}$ and 
  fluence $\Delta E_{X} \simeq 6 \times 10^{-4}$ erg
  cm$^{-2}$
 yields $t_{\rm fb} + \Delta t_{\rm offset}
        \simeq 0.9$ d for $\alpha=-5/3$, or
  $t_{\rm fb} + \Delta t_{\rm offset} \simeq 0.5$ d 
   for $\alpha=-4/3$ (Mangano et al. 2015).
     These small values $t_{\rm fb} +\Delta t_{\rm offset}
    \la 1$ d argue against the possibility for $\Delta
  t_{\rm offset} \ga 10$ d presented in Tchekhovskoy et al. (2014).

Mangano et al. (2015)
      determine a post-fluctuation 
decay slope $\alpha = -1.36 \pm 0.02$ (Figure \ref{fig:vanessa}), which
  is  statistically 
distinguishable from the $\alpha = -5/3$ value for Swift J1644+57
commonly cited in the literature. Previous studies 
quoting a slope 
 (e.g., 
Metzger, Giannios, \& Mimica 2012;
         Lei, Zhang, \& Gao  2013;
            Kawashima et al. 2013;
                Kumar et al. 2013;
      Barniol Duran \& Piran 2013;
             Zauderer et al. 2013;
             Shen \& Matzner 2014;
         Tchekhovskoy et al. 2014)
   did not carry out detailed fitting but simply
overlay a $\alpha = -5/3$ decay onto a $\log f_X - \log t$ light curve
for Swift J1644+57 taken from the \emph{Swift}/XRT archive
    (Evans et al. 2007, 2009),
which assumes a single energy-to-counts conversion factor.
   The combination of
    (i) a small inferred $t_{\rm fb}$ and 
   (ii) a decay rate $\alpha \approx -4/3$ 
   (Mangano et al. 2015)
     support the
viewpoint of a rapid transition from stellar fallback to
disk accretion, where the disk is highly advective 
  (Cannizzo, Troja, \& Lodato 2011;
     Gao 2012). A value 
$t_{\rm fb} \la 1$ d 
 challenges current theory, which favors 
  $t_{\rm fb} \simeq 20-30$ d,
but does not consider strong general 
    relativistic effects
in the Kerr metric for large $R_T/R_P$ encounters;
  modifications 
     in the spread in specific binding energy for the tidal
debris from the standard results for $R_T/R_P \simeq 1$ 
encounters are treated via linear perturbations to a Newtonian
gravitational potential 
   (e.g., Guillochon \& Ramirez-Ruiz 2013;
          Stone, Sari, \& Loeb 2013, see their Sect. 6).

\section{GRB 130603B: a possible kilonova?}

        Neutron stars represent the most compact form of matter
                 in the universe (Pines \& Alpar 1985).
  Beneath an atmosphere only a few m thick one has an outer 
crust $\sim$$0.2-0.4$ km thick,
 varying in density  from 
     $\rho\sim$$7\times 10^6  \   {\rm g}  \ {\rm cm}^{-3}$
    to 
     $\rho\sim$$4\times 10^{11} \ {\rm g}  \ {\rm cm}^{-3}$.
  It contains  a solid array of nuclei  and  highly
degenerate relativistic electron  plasma. 
The
inner crust has a radial extent of a few km
   and varies from
         $\rho\sim$$4  \times10^{11}$ ${\rm g}  \ {\rm cm}^{-3}$
      up to 
         $\rho\sim$$2  \times10^{14}$ ${\rm g}  \ {\rm cm}^{-3}$
     at its base.
    It is composed of a highly 
        degenerate superfluid neutron liquid
          in addition to a lattice of
      increasingly neutron-rich nuclei and
 relativistic electrons. 
 The quantum liquid interior is thought to be
  mainly superfluid neutrons
   with a few percent protons and electrons. 
 The density at the core  $\rho\sim$$10^{15}$ ${\rm g}  \ {\rm cm}^{-3}$;
     by  comparison 
  $\rho\sim$$2.8\times10^{14}$ ${\rm g}  \ {\rm cm}^{-3}$
  for nuclear matter.

    As discussed previously,
     the currently favored model for short GRBs
    is a NS-NS merger. During the merger 
       streamers of neutron star  material may get ejected
    and decompress.
%
%
     When the density of the expanding plasma
        falls below nuclear,
        nucleon  clusters are formed, the matter
        inside each quasi-nucleus in equilibrium 
     with the external dripped neutron sea (Lattimer et al. 1977).
  As the density drops further
these  nuclei lose neutrons to the external 
  sea of neutrons.
    When the $\beta-$decay time scale becomes shorter
   than the expansion time scale, the neutrons, which had
   previously been stable by virtue of their ``nuclear'' environment, 
                  begin to $\beta-$decay.
    The nuclei then increase in proton
  and neutron number until they become unstable
   to fission. This process will resemble 
         the standard $r-$process.
%
%

\begin{figure}[h!]
\begin{centering}
\includegraphics[width=3.20truein]{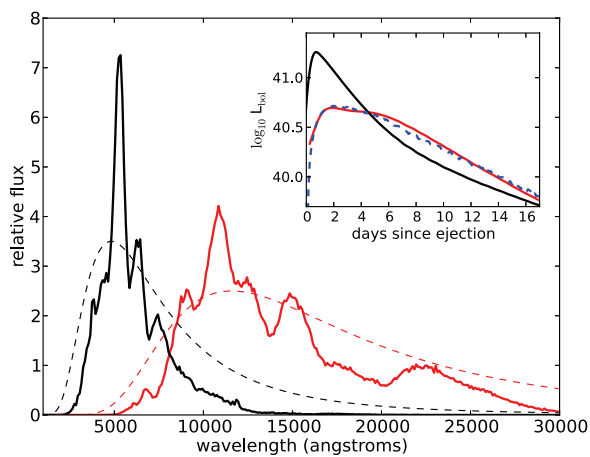}
\vskip -0.05truein
\caption{ 
   Synthetic  
        spectra (2.5 d after mass ejection) of the $r-$process SN
    model (Kasen, Badnell, \& Barnes 2013), i.e., a ``kilonova'',
          calculated using 
    Kurucz iron group opacities
   (black line) or Autostructure-derived
    $r-$process opacities (red line). For
    comparison, also shown are blackbody curves with $T = 6000$ K 
    (black
    dashed) and $T = 2500$ K (red dashed). The inset shows the corresponding
   bolometric light curves assuming iron (black) or $r-$process (red) opacities.
     Also shown is a light curve 
     calculated with a gray opacity of
    $\kappa = 10$ cm$^2$ g$^{-1}$ (blue dashed line).
  }
\label{fig:kasen}   
\end{centering}
 \end{figure}


  Various groups have explored the
   supernova-like
  transient powered by this radioactive decay 
            of material ejected from the NS
  (Eichler et al. 1989;
   Li \& Paczy\'nski 1998;
   Kulkarni 2005;
   Metzger, Piro, Quataert 2008;
   Metzger et al. 2010;
   Metzger \& Berger 2012).
    The resultant  ``kilonova''  (dimmer than a supernova and brighter than a nova)
         would
         produce 
    relatively isotropic optical/NIR
   emission after a NS-NS/NS-BH merger. 
    While SNIa  light curves are powered
 primarily by decay of $^{56}$Ni,
    the ejecta from a disrupted NS is neutron
 rich and yields little Ni.
        Much heavier radioactive
elements form via rapid neutron capture
  ($r-$process) 
  nucleosynthesis   following the decompression 
    of the ejecta from nuclear densities.
  These
newly synthesized elements undergo nuclear
 fission, $\alpha$ and $\beta$
decays on much longer time-scales. 
The resulting energy release can power detectable thermal emission
once the ejecta expands sufficiently that photons can escape.



\begin{figure}[h!]
\begin{centering}
\includegraphics[height=3.95truein]{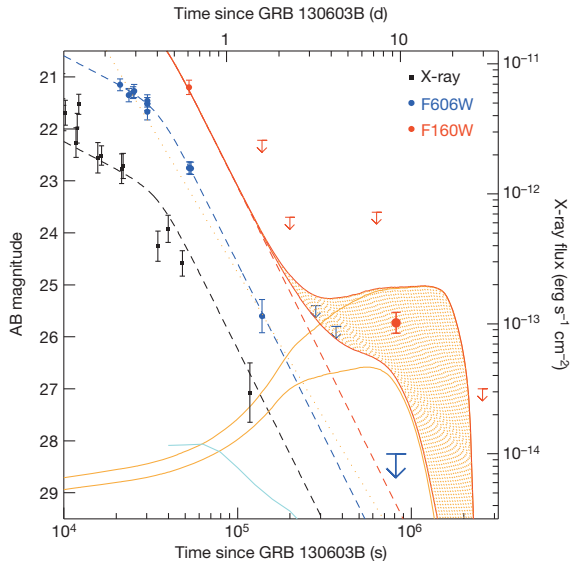}
\vskip -0.3725truein
\caption{Optical, NIR and X-ray light curves of GRB 130603B   
        (Tanvir et al. 2013). Left axis,
optical and NIR; right axis, X-ray. 
Optical data ($g$, $r$ and $i$) have been interpolated to the 
  \emph{HST} F606W band and
 NIR data have been interpolated to the F160W band
     using an average
spectral energy distribution at $\sim$0.6 d 
   Absence of late-time optical
emission places a limit on any separate $^{56}$Ni-driven 
   decay component. The
$0.3-10$ keV \emph{Swift}/XRT ray data (Evans et al. 2007, 2009) 
   are also consistent with breaking to a similarly steep
decay (the dashed black line shows the optical light curve simply rescaled to
match the X-ray points in this time frame), although the source had dropped
below \emph{Swift}/XRT
     sensitivity by $\sim$48 hr (our frame). 
   The NIR detection requires an additional component above
the extrapolation of the afterglow (red dashed line). 
   This excess NIR flux corresponds to absolute
magnitude $M(J)_{\rm AB} \approx -15.35$
     at  $\sim$7 d  (source frame). This
is consistent with the favored range of 
   kilonova behavior from recent
calculations (despite their known significant
    uncertainties:
 Kasen et al. 2013;
  Barnes  \& Kasen 2013;
   Tanaka \& Hotokezaka 2013;
   Grossman et al. 2014).   
   Model lines 
  (Barnes \& Kasen 2013; 
   orange curves) 
    correspond to ejected masses of $0.01\msun$ 
(lower curve) and $0.1\msun$ (upper curve), and these are added
to the afterglow decay curves to produce predictions for the total NIR emission,
shown as solid red curves. 
   The cyan curve shows that even the brightest
predicted $r-$process kilonova optical emission is negligible.
  }
\label{fig:tanvir}   
\end{centering}
 \end{figure}

    Kasen et al. (2013)  argue that the 
  opacity of the  expanding $r-$process material is
dominated by bound–bound transitions from those ions
    with the most complex valence electron structure, 
   i.e., 
the lanthanides. 
         They compute 
    atomic structure models for a few representative ions
         in order to calculate the radiative transition
rates for tens of millions of lines, and
     find that  resulting $r-$process 
   opacities are orders of magnitude larger than that of
ordinary (e.g., iron-rich) supernova ejecta.
The resultant light curves should be longer, dimmer, 
   and redder than  previously thought 
        (Figure \ref{fig:kasen}). The spectra 
   have broad absorption features and peak in the IR
  ($\sim$1 $\mu$m). 

What are the prospects for observing a kilonova in conjunction
    with a short GRB?  The biggest problem is that, since the
   kilonova emission
  is weak, it could usually be masked by the normal afterglow.
       The only hope is 
           for a short GRB occurring in a 
  very low density interstellar medium.
           Observational confirmation of such an event would be  important
    given that this mechanism may be the
predominant source of stable $r-$process elements in the universe
  (Freiburghaus, Rosswog, \& Thielemann 1999;
           Goriely, Bauswein, \& Janka  2011).


GRB 130603B might be the first detected 
  kilonova (Tanvir et al. 2013;
               Berger, Fong, 
                     \& Chornock 2013).
   It was a short 
          GRB at $z=0.356$ with a duration $\sim$0.2 s in the BAT.
    Tanvir et al. (2013) present
 optical and near-infrared observations that provide strong
evidence for an accompanying kilonova (Figure \ref{fig:tanvir}).
If correct, it would confirm that compact-object mergers 
  are the 
progenitors of short GRBs and also  the sites of significant
production of $r-$process elements. 
   It also offers that hope that kilonovae
           provide
            an alternative, 
    unbeamed electromagnetic signature of the
most promising sources for direct detection of gravitational waves.

\begin{figure}[h!]
\begin{centering}
\includegraphics[width=3.05truein]{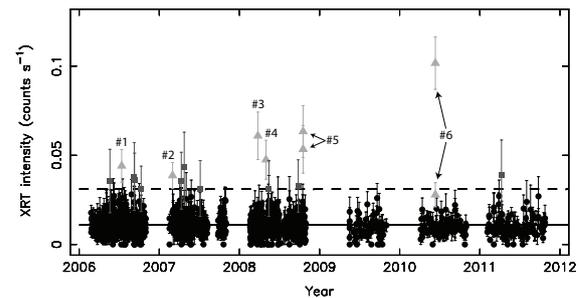}
\vskip -0.25truein
\caption{ Long term $0.3-10$ keV \emph{Swift}/XRT light curve of Sgr A$^*$ 
    (Degenaar et al. 2013). 
   %
    Solid horizontal line indicates the mean count
rate observed in 2006-2011, whereas the dashed line 
 indicates the $3\sigma$ level. The six confirmed 
    X-ray flares are numbered and indicated by
light gray triangles.
\label{fig:degenaar}   
  }
\end{centering}
 \end{figure}

\section{Sgr A$^*$ Flares}

The closest and best studied SMBH lies 
  at the heart of our galaxy, in Sgr A$^*$.
Its bolometric luminosity is lower than expected 
    from an Eddington-limited SMBH of mass
$\sim$$4\times10^6 \msun$ by a 
  factor $\sim$$10^8-10^9$, indicating the
   heyday of its quasar-like youth is well past. It has
long since depleted its ``loss cone'' 
         (Frank \& Rees 1976;
   Young, Shields, \& Wheeler 1977) 
    supply of stars and gas 
and its very low accretion rate is generally 
         characterized by a 
   radiatively inefficient accretion flow 
         (Narayan, Yi, \& Mahadevan 1995;
     Manmoto, Mineshige, \& Kusunose 1997;
                     Dexter et al. 2010;
     Shcherbakov, Penna, \& McKinney 2012).
Sgr A$^*$ emits a steady luminosity
   $\sim$$2 \times 10^{33}$ erg s$^{-1}$
   in the  soft X-ray band
 (Baganoff et al. 2003),
with occasional flaring up by a factor 
             $\sim$$5-150$ for tens of minutes to hours. 
   For $\sim$5 yr beginning
in 2006, \emph{Swift}/XRT observed a 
  $\sim$$21^{'} \times  21^{'}$ region around Sgr A$^*$. 
  Six flares  were seen, with
luminosities $\sim$$(1-3) \times 10^{35}$ erg s$^{-1}$
 (Figure \ref{fig:degenaar}). 
   Based on the number of observed flares and the
total length of observations, Degenaar et al. (2013)
   estimate a flaring rate $0.1-0.2$ d$^{-1}$. This
implies a bright flare with $L_X \simeq 10^{35}$ erg s$^{-1}$ 
    occurs every $\sim$$5-10$ d. This rate is in accord with
previous estimates based on \emph{Chandra} data (Baganoff et al. 2003).

\section{RS Oph nova}



Classical and recurrent
novae happen in interacting binaries
  containing a WD accretor
 and are due to
the thermonuclear detonation of
    accreted material on the
   surface of a WD
     (Gallagher \& Starrfield 1978).
  This can occur if the temperature
  and pressure at the base of
  the accumulated layer
  of accreted matter
    are in the appropriate regime.
    \emph{Swift} has opened a new
   window on nova studies.
%
%
  An overview of the \emph{Swift} sample of novae
 (52 galactic plus Magellanic Cloud)
   is given by Schwarz et al. (2011).
  \emph{Swift} has detected keV emission from shocked
ejecta and supersoft (SS) emission from the WD surface.

\begin{figure}[h!]
\begin{centering}
\includegraphics[width=3.15truein]{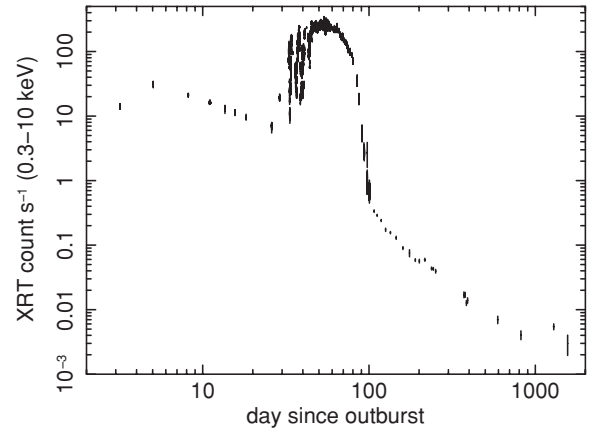}
\vskip -0.05truein
\caption{Entire $0.3-10$ keV \emph{Swift}/XRT light curve of the 2006
  outburst of the recurrent nova RS Oph (Osborne et al. 2011). 
    The supersoft phase is prominent between days 29 and 100.
  }
\label{fig:osborne}
\end{centering}
 \end{figure}

  RS Oph is a recurrent nova
   consisting of a red giant (RG) donor
  and a white dwarf (WD) accretor
     residing in a semi-detached (i.e., mass-exchange)
  binary.
   About every 20 yr
    enough material from the RG accumulates on the surface
of the WD to produce a thermonuclear explosion.
On 12 Feb 2006 a new eruption occurred, reaching $m_V \simeq 4.5$.
    Detailed analysis of \emph{Swift}
  observations  (Figure \ref{fig:osborne})
         indicated  a mass
ejection of $\sim$$3\times10^{-5}M_\odot$
  at $\sim$$4000$ km s$^{-1}$
   into the wind of the mass losing red
  giant companion in the system (Osborne et al. 2011).

Supersoft sources (SSSs) are powered 
           by residual nuclear burning
    on the WD surface following the main nova eruption.
   Once the nova shell has expanded sufficiently it
    can become optically thin to $0.2-1$ keV X-rays
 so that we can see all the way down to the WD surface
    and directly observe the nuclear reactions.



RS Oph is unusual in at least three respects compared
to most novae:
(i) The WD in the system is fairly massive so that the
    residual nuclear burning happens at a relatively
  high temperature, 
  ${T_{\rm eff}} \simeq 10^6$ K,
    and the resultant emission fills the $0.3-1$ keV bandpass.
%
%
  (ii) Its  distance is  only $\sim$2.5$ - $3 kpc
           which makes it bright.
(iii) RS Oph has a wide orbit with a red giant donor instead of a
      red dwarf. Thus it is an ``embedded nova'' because the
 shell runs into gas previously ejected in the RG wind.
         Therefore embedded novae are brighter in hard X-rays
       $\sim$$10^{36}$ erg s$^{-1}$
          than normal novae.



 Could it be that 
  SSS novae represent transitional objects between normal
   novae and Type Ia (single degenerate supernovae)?
        Not likely.  It appears that all novae have
the potential to be detected as SSS, and we have
plenty examples of SSS, all novae, within several kpc.
    However, selection effects can prevent detection,
   especially if the absorbing column is too great.
      We can see the SSS inside a nova if the absorbing column
        $ N_H \la 2\times 10^{21} \  {\rm cm}^{-2}$ 
                   (Osborne et al. 2011).






%
%
%

\section{DG CVn Superflare}

            A basic fact of stellar structure is that early spectral type
 stars O-B-A have convective cores and radiative envelopes, whereas later spectral type 
   stars   A-F-G-K-M have radiative cores and convective envelopes. 
     The dividing point lies at two solar masses 
               which corresponds roughly to an A4 star.
    Our G2 sun is convective 
 over its outermost 30\% in radius. 
                Late type M stars
                    are completely convective ($\la0.4\msun$). 
      In a subset of stars of late spectral type
  the combination of surface convection and high rotation 
    can lead to strong expulsion of magnetic fields from the
  stellar surface. 
              In stars with outer convection
         a  dynamo operates at the base of the convective
    envelope, twisting internal
dipole field into a tangled geometry.
       Magnetic buoyancy expels field from the photosphere
     as active regions.  
             Loops arch outward from 
   the stellar surface,  extending from ``$-$'' to ``$+$'' polarity.
          Where two loops cross one can have a massive reconnection
   event -- a superflare.  
 
 By contrast, although early spectral type stars such as Ap and Am can have 
   strong magnetic fields,  
   they are convective only at their cores
    and therefore
       do not actively transport B-field to outside the star.
Thus they do not have stellar flares. 
   Their high fields are only evidenced through Zeeman splitting
  of photospheric spectral lines.

 DG CVn, a close visual  dM4e$+$dM4e 
    binary with orbital separation $\sim$3 AU, 
      is a known flare star system. 
          Both stars have
masses and radii $\sim$1/3 solar.
             At 18 pc the system 
                           is relatively close. 
     Its kinematics identify it as being young, $\sim$30 Myr, 
    and furthermore at least one of the stars  
             is a fast rotator with $v\sin i \simeq 50$ km s$^{-1}$.
 For comparison, if our sun were
     examined spectroscopically  
     from several pc
    at a random
        orientation it would have $v\sin i \approx 1-2$ km s$^{-1}$.

\begin{figure}[h!]
\begin{centering}
\includegraphics[width=3.20truein]{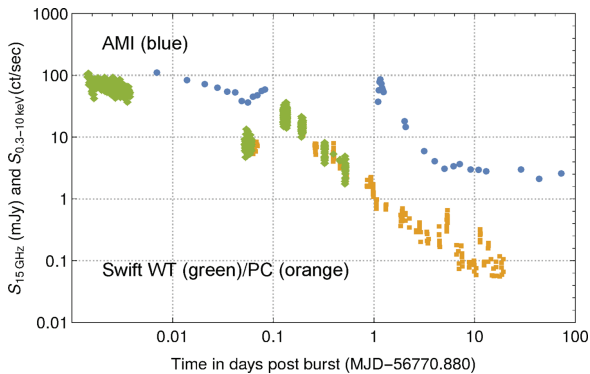}
\vskip -0.05truein
\caption{ 
Superflare from the DG Cvn (Fender et al. 2015).
  Light curves are from
AMI-LA $13-18$ GHz (blue circles) and \emph{Swift}/XRT  WT/PC 
  (green diamonds/orange squares) $0.3-10$ keV. 
In X-rays the source was brightest
       at the first measurement,
   $T0 + 2$ min.,
      and then declined for $\sim$1 hr,
   rebrightening
 at  $0.075  <  (t-T0)<    0.125$ d. 
     The radio flux behaved similarly, 
   with a strong detection 
  in the first measurement at $T0 +  6$ min.,
     followed by a decline
and subsequent rebrightening. A second
   radio 
   flare  
  occurred at $\sim$$T0 + 1.1$ d. 
    }
\label{fig:fender}   
\end{centering}
 \end{figure}


%










On 23 April 2014 \emph{Swift}/BAT detected a superflare  in 
DG CVn   which  
   reached 0.3 Crab in the $15-150$ keV BAT band.
The flare  arose from one of the stars in the binary.  
%
    It consisted of a series of
outbursts; the strongest was $\sim$$10^4$ times more energetic
    than the largest solar flare ever seen -- the
Carrington Event of 1859.
Time resolved spectral fitting at the peak of the flare
implies $T \simeq  2  \times  10^8$ K and
     $L_X \simeq 1.9 \times 10^{32}$
    erg s$^{-1}$ within the XRT $0.3-10$ keV window.
    This compares with a normal systemic bolometric 
   luminosity $1.3 \times  10^{32}$ erg s$^{-1}$.
As with a previous superflare seen in 2008 in EV Lac (Osten et al. 2010), 
 for several minutes the
X-ray emission from the flare outshone the total light from the system. 
 %


   Alerted by the \emph{Swift}/BAT trigger,
  Fender et al. (2015)    
              detected a bright ($\sim$100 mJy) radio 
     flare using AMI-LA (Figure \ref{fig:fender}).
   This is the earliest detection
   ever made
    of 
bright, prompt, radio emission from a high-energy 
  transient.   
Although radio emission is known to be associated with 
   active stars, this was the first detection
of a large radio flare 
      in conjunction with a gamma-ray superflare.
\section{Conclusion}

 The last ten years have been a time of
    great discovery  for 
\emph{Swift}.
   The
    sky is rich in transients of many types,
   and
\emph{Swift} is exploring the transient sky with 
   unprecedented sensitivity and coverage.
   Every year brings a new discovery in time domain science.
    Explosion mechanisms range from gravitational collapse
        to nuclear burning 
        to \emph{B} field reconnection.
   \emph{Swift}
    will hopefully last another ten years and
    have exciting science to perform 
  while partnering with
   new ground observatories of increasing capabilities.





\begin{thebibliography}{86}

\bibitem{1a}
  Antonelli L.A.,
    et al. 
 2005. GCN, 3678, 1.

\bibitem{a2}
  Baganoff F.K., 
  et al., 
     2003. ApJ, 591, 891.


\bibitem{4a}
   Barnes J.,
    Kasen D.,
   2013. ApJ, 775, 18.

\bibitem{5}
  Barniol Duran R.,
          Piran T.,
  2013. ApJ, 770, 146.

\bibitem{6}
  Barthelmy S.D., 
  et al., 
     2005. Nature, 438, 994.


\bibitem{8berg}
   Berger E.,
   2014. ARA\&A, 52, 43.

\bibitem{8}
  Berger E., 
   et al., 
   2005a. ApJ, 634, 501.

\bibitem{9}
  Berger E. 
     et al., 
  2005b. Nature, 438, 988.

\bibitem{7}
        Berger E.,     
      Zauderer A.,     
        Pooley G.G.,   
     Soderberg A.M.,   
          Sari R.,     
    Brunthaler A.,     
    Bietenholz M.F.,   
   2012. ApJ, 748, 36.

\bibitem{10ber}
        Berger E.,
          Fong W., 
      Chornock R.,
  2013. ApJ, 774, L23.

\bibitem{10}
  Berger E., 
   et al., 
   2014. ApJ, 796, 96.



\bibitem{11}
   Bloom J.S., et al., 
    2006. ApJ, 638, 354.

\bibitem{12}
    Bloom J.S.,  
    et al.,
 2009. ApJ, 691, 723.	

\bibitem{13}
   Bloom J.S., et al.,
   2011. Science, 333, 203.

\bibitem{14}
   Burrows D.N., et al., 
   2005. Science, 309, 1833.

\bibitem{15}
   Burrows D.N.,
    et al.,
   2011. Nature, 476, 421.

\bibitem{c15}
    Campana S., 
   et al.,
  2006. Nature, 442, 1008.

\bibitem{15campa}
    Campana S., 
     et al.,
  2012. MNRAS, 421, 1697.

\bibitem{15cg}
   Cannizzo J.K.,
    Gehrels N.,
   2009. ApJ, 700, 1047.

\bibitem{16}
    Cannizzo J.K.,
         Lee H.M., 
     Goodman J.,
   1990. ApJ, 351, 38.

\bibitem{17}
     Cannizzo J.K., 
        Troja E., 
       Lodato G.,
   2011. ApJ, 742, 32.

\bibitem{18}
  Castro-Tirado A.J., 
    et al.,  
   2005. A\&A, 439, L15.

\bibitem{18che}
    Chevalier  R.A.,
           Li Z.-Y.,
     2000. ApJ, 536, 195.

\bibitem{18chin}
    Chincarini G.,
    et al.,
     2007. ApJ, 671, 1903.

\bibitem{19}
   Cucchiara A.,  
     et al.,
     2011. ApJ, 736, 7.

\bibitem{19da}
    Dall'Osso S.,
      Stratta G.,
       Guetta D.,
       Covino S.,
    De Cesare G.,
       Stella L.,	
  2011. A\&A, 526, A121.

\bibitem{20}
     D'Avanzo P.,
      et al.,
  2005. GCN, 3690, 1.

\bibitem{20davanz}
     D'Avanzo P.,
      et al.,
  2012. MNRAS, 425, 506.

\bibitem{21}
   Degenaar N., 
     Miller J.M., 
     Kennea J.,
    Gehrels N.,
   Reynolds M.T., 
   Wijnands R.,
     2013. ApJ, 769, 155.

\bibitem{22della}
  Della Valle M.,
  et al.,
   2006. Nature, 444, 1050


\bibitem{22d}
  de Ugarte Postigo A.,
  et al.,
 2013. CBET, 3529, 1.

\bibitem{22}
      Dexter J.,
        Agol E.,
     Fragile P.C.,
    McKinney J.C.,
     2010. ApJ, 717, 1092.

\bibitem{22d}
      Duffell P.C.,
    MacFadyen A.I.,
   2014. astro-ph/1407.8250

\bibitem{23}
     Eichler D.,
       Livio M., 
       Piran T.,
     Schramm D.N.,
     1989. Nature, 340, 126.

\bibitem{24}
  Evans P.A., 
   et al.,
   2007. A\&A, 469, 379.

\bibitem{25}
   Evans P.A., 
     et al.,
   2009. MNRAS, 397, 1177.

\bibitem{25falcone}
   Falcone A.D., 
     et al.,
   2007. ApJ, 671, 1921.

\bibitem{26}
     Fender R.P.,     
   Anderson G.E.,   
      Osten R.,        
     Staley T.,       
     Rumsey C.,       
    Grainge K.,       
   Saunders R.D.E.,  
   2015. MNRAS, 446, L66.

\bibitem{fb27}
     Fong W.,
   Berger E.,
   2013. ApJ, 776, 18.

\bibitem{f27}
     Fong W.,
   Berger E.,
      Fox D.B.,
   2010. ApJ, 708, 9.

\bibitem{27}
 Fox D.B., 
   et al.,  
 2005. Nature, 437, 845.

\bibitem{28}
    Frank J.,
     Rees M.J.,
    1976. MNRAS, 176, 633.

\bibitem{29}
  Freiburghaus C., 
       Rosswog S.,
    Thielemann F.-K.,
  1999. ApJ, 525, L121. 

\bibitem{29fynbo}
  Fynbo J.P.U.,
      et al.,
 2006. Nature, 444, 1047.

\bibitem{30}
     Gallagher J.S., 
    Starrfield S.,
   1978. ARA\&A, 16, 171.

\bibitem{31}
   Gal-Yam A.,
     et al.,
  2005. GCN, 3681, 1.

\bibitem{32}
  Gao W.-H.,
   2012. ApJ, 761, 113.

\bibitem{33}
 Gehrels N., 
   et al.,
 2004. ApJ, 611, 1005.

\bibitem{34}
Gehrels N.,
    et al., 
 2005. Nature, 437, 851.

\bibitem{34gehrels2006}
Gehrels N.,
    et al., 
 2006. Nature, 444, 1044.

\bibitem{34gehr}
       Gehrels N.,
  Ramirez-Ruiz E.,
           Fox D.B.,
    2009. ARA\&A, 47, 567.

\bibitem{35}
  Gezari S.,  
     et al.,
   2006. ApJ, 653, L25.

\bibitem{36}
   Gezari S.,  
    et al.,
   2008. ApJ, 676, 944.

\bibitem{36ghirla}
   Ghirlanda G.,
    et al.,
   2013. MNRAS, 435, 2543.

\bibitem{37}
    Goriely S., 
   Bauswein A.,
      Janka H.-T.,
  2011. ApJ, 738, L32.

\bibitem{38}
     Grossman D.,
     Korobkin O.,
      Rosswog S.,
        Piran T.,
    2014. MNRAS, 439, 757.


\bibitem{38grupe}
       Grupe D.,
     Burrows D.N.,
       Patel S.K.,
 Kouveliotou C.,
       Zhang B.,
  M\'esz\'aros P.,
   Wijers R.A.M.,
    Gehrels N.,
  2006. ApJ, 653, 462.

\bibitem{39}
        Guillochon J.,
      Ramirez-Ruiz E.,
      2013. ApJ, 767, 25.


\bibitem{40h}
  Hjorth J., 
      et al.,
  2003. Nature, 423, 847.

\bibitem{40}
  Hjorth J., 
     et al., 
  2005a. ApJ, 630, L117.

\bibitem{41}
    Hjorth J., 
   et al., 
   2005b. Nature, 437, 859.

\bibitem{41hjorth2012}
    Hjorth J.,
   et al.,
   2012. ApJ, 756, 187.

\bibitem{41jak}
  Jakobsson P.,
   et al.,
  2012. ApJ, 752, 62.

\bibitem{42}
    Kasen D.,
  Badnell N.R.,
   Barnes J.,
   2013. ApJ, 774, 25.

\bibitem{43}
     Kawashima T.,
        Ohsuga K.,
          Usui R.,
         Kawai N.,
        Negoro H.,
     Matsumoto R.,
   2013. PASJ, 65, L8.

\bibitem{44}
  Kennicutt Jr R.C.,
   1998. ARA\&A, 36, 189.


\bibitem{45}
   Kobayashi S., 
       Zhang B.,
  2003a. ApJ, 582, L75.

\bibitem{46}
   Kobayashi S., 
       Zhang B.,
   2003b. ApJ, 597, 455.

\bibitem{46kohn}
   Kohn S.A., 
    et al.,
   2015. MNRAS, 448, 1494.

\bibitem{47}
        Komossa S.,  
        Halpern J.,  
       Schartel N.,  
       Hasinger G.,  
    Santos-Lleo M.,  
        Predehl P.,  
    2004. ApJ, 603, L17.

\bibitem{48kruehl}
   Kr\"uhler T.,
      et al.,
   2012. ApJ, 758, 46.

\bibitem{48kul}
   Kulkarni S.R.,
  2005. astro-ph/0510256

\bibitem{48k}
      Kumar P.,
 Panaitescu A.,
    2000. ApJ, 541, L51.

\bibitem{48k1}
      Kumar P.,
    Narayan R.,
    Johnson J.L.,
   2008. MNRAS, 388, 1729.

\bibitem{48}
             Kumar P., 
     Barniol Duran R.,
       Bo\v{s}njak \v{Z}.,
             Piran T.,
   2013. MNRAS, 434, 3078.


\bibitem{49}
  Laskar T., 
     et al., 
  2013. ApJ, 776, 119.  

\bibitem{50}
     Lattimer J.M.,
       Mackie F.,
    Ravenhall D.G.,
      Schramm D.N.,
      1977. ApJ, 213, 225.

\bibitem{51}
       Lei W.-H.,
     Zhang B.,
       Gao H.,
   2013. ApJ, 762, 98.

\bibitem{52}  
    Levan A.J.,
       et al.,
   2011. Science, 333, 199.

\bibitem{53}
      Levan A.J., 
      Cenko S.B.,
     Perley D.A.,
     Tanvir N.R.,
  2013. GCN, 14455, 1.

\bibitem{54}
  Levan A.J., 
      et al., 
   2014. ApJ, 792, 115.

\bibitem{55} 
             Li L.-X.,
    Paczy\'nski B.,
     1998. ApJ, 507, L59.

\bibitem{56}
          Maiolino R., 
         Schneider R., 
             Oliva E., 
           Bianchi S., 
           Ferrara A., 
          Mannucci F., 
            Pedani M., 
       Roca Sogorb M., 
      2004. Nature, 431, 533.

\bibitem{57}
       Mangano V., 
       Burrows D.N.,
    Sbarufatti B.,
      Cannizzo J.K.,
         2015. ApJ, submitted. 

\bibitem{58}
    Manmoto T.,
  Mineshige S.,
   Kusunose M.,
   1997. ApJ, 489, 791.

\bibitem{58masel}
    Maselli A.,
     et al.,
      2014. Science, 343, 48.

\bibitem{m59}
   Mazzali P.A., 
  et al.,
2006. Nature, 442, 1018.

\bibitem{mazz59pl}
   Mazzali P.A.,
  et al.,
2008. Science, 321, 1185.

\bibitem{58melandri}
   Melandri A.,
    et al.,
   2014. A\&A, 567, A29.

\bibitem{59}
   Metzger B.D.,
    Berger E.,
   2012. ApJ, 746, 48.

\bibitem{62}
   Metzger B.D.,
      Piro A.L.,
  Quataert E.,
   2008. MNRAS, 390, 781.

\bibitem{61}
   Metzger B.D., 
    et al.,
  2010. MNRAS, 406, 2650.

\bibitem{60}
    Metzger B.D.,
   Giannios D.,
     Mimica P.,
   2012. MNRAS, 420, 3528.

\bibitem{63michal}
  Micha\l{}owski M.J.,
     et al.,
   2012. ApJ, 755, 85.

\bibitem{63}
   Micha\l{}owski M.J., 
     et al.,
        2014. A\&A, 562, A70.

\bibitem{63milvang}
  Milvang-Jensen B.,
           Fynbo J.P.U.,
        Malesani D.,
          Hjorth J.,
       Jakobsson P.,
       M{\o}ller P.,
    2012. ApJ, 756, 25.

\bibitem{64}
     Nousek J.A.,
       et al., 
      2006. ApJ, 642, 389.

\bibitem{64o}
    O'Brien  P.T.,  
    et al., 
    2006. ApJ, 647, 1213.

 \bibitem{64obr}
     O'Brien    P.T., 
   Rowlinson      A.,
   2012. 
    in IAU Symp. 279, 
       Death of Massive
           Stars: Supernovae and Gamma-Ray Bursts, 
 ed. P.W.A. Roming, N. Kawai,
     \& E. Pian,
   (Cambridge: Cambridge Univ. Press), 297.

\bibitem{65}
     Osborne J.P., 
      et al., 
   2011. ApJ, 727, 124.

\bibitem{66}
    Osten R.A., 
       et al.,
    2010. ApJ, 721, 785.

\bibitem{67}
     Ota K.,
      et al., 
    2014. ApJ, 792, 34.

\bibitem{68}
   Ouchi M., 
        et al.,
   2013. ApJ, 778, 102.

\bibitem{69}
    Narayan R.,
         Yi I.,
  Mahadevan R.,
  1995. Nature, 374, 623.

\bibitem{69nava}
     Nava L.,
       et al.,
  2012. MNRAS, 421, 1256.

\bibitem{70}
  Perley D.A., 
  et al., 
  2014. ApJ, 781, 37.

\bibitem{71}
  Phinney E.S.,
  1989.
    in IAU Symp. 136,  The Center of the Galaxy,
 ed. M. Morris (Dordrecht: Kluwer), 543.

\bibitem{p72}
   Pian E., 
   et al.,
 2006. Nature, 442, 1011.

\bibitem{72}
   Pines D.,
   Alpar M.A.,
   1985. Nature 316, 27.

\bibitem{73}
Prochaska J.X.,
     et al.,
  2005. GCN, 3700, 1.

\bibitem{74}
 Racusin J.L., 
  et al., 
 2008. Nature, 455, 183.

\bibitem{75}
 Rees M.J.,
 1988. Nature, 333, 523.

\bibitem{76}
    Ricker G.R.,
      et al., 
 2003. 
     in AIP Conf. Ser. 662, Gamma-Ray Burst and Afterglow Astronomy 2001: A Workshop
   Celebrating the First Year of the \emph{HETE} Mission, ed. G.R. Ricker
  \& R.K. Vanderspek (Mellville: AIP), 3.

\bibitem{77}
                 Rieke G.H.,  
        Alonso-Herrero   A.,  
                Weiner B.J.,  
    P\'erez-Gonz\'alez P.G.,  
              Blaylock   M.,  
                Donley J.L.,  
             Marcillac   D.,  
         2009. ApJ, 692, 556.

\bibitem{77r}
   Rowlinson   A.,
     O'Brien P.T.,
     Metzger B.D., 
      Tanvir N.R., 
       Levan A.J.,
       2013. MNRAS, 430, 1061.

\bibitem{78salva}
   Salvaterra R.,
  et al., 
   2009. Nature, 461, 1258.

\bibitem{78salvaterr}
   Salvaterra R.,
  et al., 
   2012. ApJ, 749, 68.

\bibitem{78}
    Sari R., 
   Piran T.,
   1999a. ApJ, 517, L109.

\bibitem{79}
   Sari R., 
  Piran T.,
   1999b. ApJ, 520, 641. 

\bibitem{77s}
    Sari R.,
   Piran T.,
 Narayan R.,
   1998. ApJ, 497, L17.

\bibitem{80}
Schwarz G.J., 
   et al., 
 2011. ApJS, 197, 31.

\bibitem{81}
   Shcherbakov R.V.,
         Penna R.F.,
      McKinney J.C.,
     2012. ApJ, 755, 133.

\bibitem{82}
      Shen R.-F.,
   Matzner C.D.,
    2014. ApJ, 784, 87.

\bibitem{s82}
  Soderberg A.M., 
     et al.,
  2006. Nature, 442, 1014.

\bibitem{84}
  Soderberg A.M.,
   Chandra P.,
     Frail D.,    
  2008. GCN, 7506, 1.
%
%
%
%

\bibitem{83}
   Soderberg A.M.,
        et al.,
   2008. Nature, 453, 469.  

\bibitem{85} 
     Stone N., 
      Sari R., 
      Loeb A., 
  2013. MNRAS, 435, 1809.

\bibitem{86tagli}
      Tagliaferri G.,
       et al.,
     2005. Nature, 436, 985.

\bibitem{86}
      Tanaka M.,
  Hotokezaka K.,
   2013. ApJ, 775, 113.

\bibitem{87tan}
      Tanvir N.R.,  
       et al., 
  2009. Nature, 461, 1254. 

\bibitem{87}
      Tanvir N.R.,   
       Levan A.J.,   
    Fruchter A.S.,   
      Hjorth J.,   
    Hounsell R.A.,   
    Wiersema K.,    
 Tunnicliffe R.L.,   
   2013. Nature, 500, 547.
 
\bibitem{88}
   Tchekhovskoy A.,
        Metzger B.D., 
       Giannios D.,
         Kelley L.Z.,
    2014. MNRAS, 437, 2744.

\bibitem{v88}
   Villasenor J.S., 
     et al., 
    2005. Nature, 437, 855.

\bibitem{89}
    Wang W.-H.,
    Chen H.-W.,
   Huang K.-Y.,
 2012. ApJ, 761, L32.

\bibitem{9a0}
  Xu D., 
  et al.,
  2013. ApJ, 776, 98.

\bibitem{90yam}
   Yamazaki R.
   2009. ApJ, 690, L118.

\bibitem{91}
     Young P.J.,
   Shields G.A.,
   Wheeler J.C.,
  1977. ApJ, 212, 367.


\bibitem{93}
       Zauderer B.A.,
         Berger E.,
       Margutti R.,
         Pooley G.G.,
           Sari R.,
      Soderberg A.M.,
     Brunthaler A.,
     Bietenholz M.F.,
   2013. ApJ, 767, 152

\bibitem{a94} 
          Zhang B., 
            Fan Y.Z., 
           Dyks J.,
      Kobayashi S.,
   M\'esz\'aros P., 
        Burrows D.N.,
         Nousek J.A.,
        Gehrels N.,
  2006. ApJ, 642, 354.

\bibitem{a95}
   Zou Y.C., 
    Wu X.F.,
   Dai Z.G.,
   2005. MNRAS, 363, 93.

\end{thebibliography}






\end{document}